\documentstyle[psfig]{mn}

\newif\ifAMStwofonts

\ifoldfss
  \ifCUPmtlplainloaded \else
    \NewTextAlphabet{textbfit} {cmbxti10} {}
    \NewTextAlphabet{textbfss} {cmssbx10} {}
    \NewMathAlphabet{mathbfit} {cmbxti10} {} 
    \NewMathAlphabet{mathbfss} {cmssbx10} {} 
  \fi
  \ifAMStwofonts
    \ifCUPmtlplainloaded \else
      \NewSymbolFont{upmath} {eurm10}
      \NewSymbolFont{AMSa} {msam10}
      \NewMathSymbol{\upi}     {0}{upmath}{19}
      \NewMathSymbol{\umu}     {0}{upmath}{16}
      \NewMathSymbol{\upartial}{0}{upmath}{40}
      \NewMathSymbol{\leqslant}{3}{AMSa}{36}
      \NewMathSymbol{\geqslant}{3}{AMSa}{3E}

       \let\le=\leqslant
       
    \fi
  \fi
\fi 

\ifnfssone
  \newmathalphabet{\mathit}
  \addtoversion{normal}{\mathit}{cmr}{m}{it}
  \addtoversion{bold}{\mathit}{cmr}{bx}{it}
  \newmathalphabet{\mathbfit} 
  \addtoversion{normal}{\mathbfit}{cmr}{bx}{it}
  \addtoversion{bold}{\mathbfit}{cmr}{bx}{it}
  \newmathalphabet{\mathbfss} 
  \addtoversion{normal}{\mathbfss}{cmss}{bx}{n}
  \addtoversion{bold}{\mathbfss}{cmss}{bx}{n}
  \ifAMStwofonts
    \ifCUPmtlplainloaded \else
      \UseAMStwoboldmath
      \makeatletter
      \new@mathgroup\upmath@group
      \define@mathgroup\mv@normal\upmath@group{eur}{m}{n}
      \define@mathgroup\mv@bold\upmath@group{eur}{b}{n}
      \edef\UPM{\hexnumber\upmath@group}
      \new@mathgroup\amsa@group
      \define@mathgroup\mv@normal\amsa@group{msa}{m}{n}
      \define@mathgroup\mv@bold\amsa@group{msa}{m}{n}
      \edef\AMSa{\hexnumber\amsa@group}
      \makeatother
      \mathchardef\upi="0\UPM19
      \mathchardef\umu="0\UPM16
      \mathchardef\upartial="0\UPM40
      \mathchardef\leqslant="3\AMSa36
      \mathchardef\geqslant="3\AMSa3E

       \let\le=\leqslant

    \fi
  \fi
\fi 

\ifnfsstwo
  \DeclareMathAlphabet{\mathbfit}{OT1}{cmr}{bx}{it}
  \SetMathAlphabet\mathbfit{bold}{OT1}{cmr}{bx}{it}
  \DeclareMathAlphabet{\mathbfss}{OT1}{cmss}{bx}{n}
  \SetMathAlphabet\mathbfss{bold}{OT1}{cmss}{bx}{n}
  \ifAMStwofonts
    \ifCUPmtlplainloaded \else
      \DeclareSymbolFont{UPM}{U}{eur}{m}{n}
      \SetSymbolFont{UPM}{bold}{U}{eur}{b}{n}
      \DeclareSymbolFont{AMSa}{U}{msa}{m}{n}
      \DeclareMathSymbol{\upi}{0}{UPM}{"19}
      \DeclareMathSymbol{\umu}{0}{UPM}{"16}
      \DeclareMathSymbol{\upartial}{0}{UPM}{"40}
      \DeclareMathSymbol{\leqslant}{3}{AMSa}{"36}
      \DeclareMathSymbol{\geqslant}{3}{AMSa}{"3E}

       \let\le=\leqslant

    \fi
  \fi
\fi 

\ifCUPmtlplainloaded \else
  \ifAMStwofonts \else 
    \def\upi{\pi}
    \def\umu{\mu}
    \def\upartial{\partial}
  \fi
\fi

\title{On the pulsation parallax of the variable star RR Lyr}
\author[G. Bono et al.]
 {G.Bono$^1$, F.Caputo$^1$, V.Castellani$^{1,2}$, M. Marconi$^3$, J. Storm$^4$\\
   $^1$ INAF-Osservatorio Astronomico di Roma, via di Frascati 33, 00040 Monte
        Porzio Catone, Italy\\bono/caputo/vittorio@mporzio.astro.it\\
   $^2$ INFN - Sezione di Ferrara, \\
   $^3$ INAF-Osservatorio Astronomico di Capodimonte, via Moiariello 16, 80131
        Napoli, Italy~~marcella@na.astro.it\\
   $^4$ Astrophysikalisches Institut Potsdam, An der Sternwarte 16, 
        14482 Potsdam, Germany~~jstorm@aip.de}

\date{}

\begin{document}

\maketitle

\label{firstpage}

\begin{abstract}
We show that a straightforward application of the predicted  
relation connecting the absolute $K$ magnitude of fundamental 
RR Lyrae variables to their period and metal content, together with 
current evolutionary predictions on the mass and luminosity of Horizontal 
Branch stars,  supply a distance estimate to the prototype star RR Lyr 
in close agreement with the recent parallax determination by HST 
(Benedict et al. 2002), largely unaffected by  the interstellar 
extinction correction. Adopting $<A_V>$=0.12$\pm$0.10 mag as a safe 
estimate of the extinction  correction to RR Lyr, we obtain a 
"pulsation" parallax $\pi_{puls}$=3.858$\pm$0.131 mas which 
agrees quite well, and with a smaller formal error,  
with the HST measurement ($\pi_{abs}$=3.82$\pm$0.20 mas) 
and with the weighted average of HST, Hipparcos, and ground-based 
determinations ($<\pi_{abs}>$=3.87$\pm$0.19 mas). This result suggests that 
near-infrared magnitudes and pulsation models could  
constrain the distance to well-studied RR Lyrae stars, both in the 
field and in globular clusters, with an accuracy better than current 
direct trigonometric measurements. 
\end{abstract} 


\section{Introduction}

The intrinsic luminosity of RR Lyrae variables and its dependence on 
metal content are widely debated
issues in the recent literature, mainly because of the relevance in
determining the distance to globular clusters, and in turn the age 
of these old primeval stellar systems.  Unfortunately, 
neither empirical nor theoretical approaches have provided 
firm estimates of the absolute magnitude of RR Lyrae stars yet 
(see Caputo et al. 2000, and references therein), nor   
the trigonometric parallaxes measured by HIPPARCOS for a limited 
sample of field RR Lyrae stars are precise enough to put sound 
constraints on such a parameter (see  Cassisi et al. 1998; 
Fernley et al. 1998, hereinafter F98; Groenewegen \& Salaris 1999).  
In this context, new and accurate data on the distance to these 
variables appear of great relevance and could help to establish 
the absolute distance scale of RR Lyrae stars. 

In a recent investigation, Benedict et al. (2002, hereinafter B02) 
use new astrometric data collected with FGS~3, the interferometer on 
board the Hubble Space Telescope (HST), to provide an accurate 
estimate of the absolute trigonometric parallax of the variable star 
RR Lyr, namely $\pi_{abs}$=3.82$\pm$0.20 mas. 
The uncertainty affecting this estimate is approximately a factor 
of three smaller than the uncertainty of the HIPPARCOS parallax 
($\pi_{abs}$=4.38$\pm$0.59 mas; Perryman et al. 1997), and 
dramatically smaller than the error affecting the average ground-based 
measurement ($\pi_{abs}$=3.0$\pm$1.9 mas; van Altena, Lee, \& Hoffleit 1995,
hereinafter YPC5). According to the HST parallax, one derives for 
RR Lyr a 
true distance modulus $\mu_0$=7.090$\pm$0.114 mag, while the 
weighted average of HST, Hipparcos, and YPC5 results 
($<\pi_{abs}>$=3.87$\pm$0.19 mas) yields $\mu_0$=7.061$\pm$0.107 mag. 

In this letter we show that a straightforward application of
theoretical constraints concerning the predicted $K$-band magnitude 
of fundamental RR Lyrae stars does provide a distance
to RR Lyr which is in close agreement with the HST result, and with  
a smaller formal error.

\section{$K$-band parallax}

It has already been suggested that several problems affecting
the RR Lyrae distance scale can be overcome using $K$-band
magnitudes. In a seminal empirical investigation, 
Longmore et al. (1990) found that RR Lyrae stars do 
obey a tight Period-Luminosity relation ($PL_K$) in this band.
Moreover, $K$-band data present additional advantages when
compared with optical ones: they are marginally affected by
uncertainties on reddening and present a smaller dependence on metal
content, as well as a smaller luminosity amplitude. These
interesting features were soundly enriched by the empirical
evidence brought forward by Longmore et al. (1990) that the 
slope of the $PL_K$ relation is, within
observational errors, quite constant when moving from metal-poor
to metal-rich cluster RR Lyrae stars.

These findings  have  recently  found  theoretical support. Based
on a wide grid of nonlinear, convective pulsation models, and
using bolometric corrections and color-temperature transformations
by Castelli, Gratton \& Kurucz (1997a,b), we predicted (Bono et
al. 2001, Paper I) that RR Lyrae stars obey a very tight
relation connecting the absolute $K$-band magnitude, period, 
intrinsic luminosity, and metal abundance, i.e. a $PLZ_K$
relation. This relation is characterized by a mild  
dependence on stellar luminosity and  
metal content.

We have implemented this theoretical framework by computing new sequences 
of fundamental (F) and first-overtone (FO) models that cover in 
great detail the range of metallicity  ($Z$=0.0001 to $Z$=0.02), 
luminosity ($\log L/L_\odot \sim$ 1.5 to 1.8), and mass 
($M/M_\odot \sim$ 0.5 to 0.8), as expected for field and globular cluster 
RR Lyrae stars,  
adopting a helium-to-metal enrichment ratio $\Delta Y/\Delta Z \sim $ 2.5.
The full set of F and FO models will be presented
and discussed in a forthcoming paper (Bono et al. 2002, Paper III). 
Here we report that, 
according to this theoretical scenario, the predicted absolute 
$K$ magnitude of F pulsators is correlated with the period ($P$ in days), 
metal content ($Z$), stellar mass ($M$ in solar units) 
and luminosity ($L$ in solar unitis) as follows: 

$$M_K= 0.511-2.102\log P+0.095\log Z-0.734\log L-1.735\log M/M_r\eqno(1)$$

\noindent
with an intrinsic standard deviation of 
$\sigma_K$=0.016 mag. The values of the reference mass $M_r$ for 
the various assumptions on the metallicity are given in Table 1, as  
chosen to roughly account for the correlation between metal 
content and mass of Zero Age Horizontal Branch (ZAHB) models populating 
the RR Lyrae instability strip, i.e. for effective temperatures ranging 
from 5900 to 7100 K. For the sake of comparison, Table 1 also lists 
the evolutionary mass (with an average uncertainty $\sim$ 2\%) of ZAHB models with 
log$T_e$=3.85 and 3.80, as predicted by evolutionary computations available 
in the recent literature. 

In agreement with the conclusions presented in Paper I, we find that the 
predicted fundamental $PLZ_K$ relation presents a quite small dependence on 
stellar luminosity (a variation of 0.1 dex in luminosity yields 
$\delta M_K\sim$ 0.07 mag), and metal abundance (a variation of 0.3 dex 
in metallicity yields $\delta M_K\sim$ 0.03 mag). This evidence suggests  
that the near-infrared magnitudes of RR Lyrae stars could be excellent standard 
candles to estimate the distance to field and globular cluster variables.

\begin{table}
\center \caption[]{For each adopted metal content, we list the reference 
mass $M_r$ and the evolutionary mass of ZAHB models with 
log$T_e$=3.85 and 3.80. The mass values are in solar units.\label{tab1}}

\begin{tabular}{llcc}
 $Z$ & $M_r$ & $M(3.85)$ & $M(3.80)$\\
 0.0001 & 0.75 & 0.796 & 0.852\\
 0.0004 & 0.70 & 0.699 & 0.721\\
 0.001  & 0.65 & 0.648 & 0.666\\
 0.006  & 0.58 & 0.585 & 0.589\\
 0.01   & 0.58 & 0.575 & 0.578\\
 0.02   & 0.53 & 0.545 & 0.546\\
\end{tabular}
\end{table}

We can now use the predicted $PLZ_K$ relation together with 
the observed near-infrared magnitude $K=6.54\pm0.04$ mag (Fernley, 
Skillen \& Burki 1993) of RR Lyr to derive a {\it pulsation} parallax 
to be compared with the HST trigonometric measurement.  

According to B02, the interstellar extinction toward 
RR Lyr is $<A_V>$=0.07$\pm$0.03 mag. Since $A_K\sim$ 0.11 $A_V$
(Cardelli et al. 1989), this yields that the infrared extinction correction  
is at most of the 
order of 0.01 mag and therefore it can be neglected. Thus, on the 
basis of the HST true distance modulus (7.090$\pm$0.114 mag, B02) 
one finds $M_K$=$-$0.550$\pm$0.118 mag, where the error is practically 
given by the uncertainty on the absolute trigonometric parallax. 

On the theoretical side, the predicted $PLZ_K$ relation [eq. (1)]
provides an independent estimate of the near-infrared absolute
magnitude, once the metallicity, the mass, and the luminosity 
of the variable are known. As for the metallicity of RR Lyr, 
the measured iron-to-hydrogen ratio [Fe/H]=$-$1.39 (F98) implies 
a metal abundance $Z\sim$ 0.0008, or slightly larger if the
$\alpha$-elements are overabundant with respect to iron 
(Clementini et al. 1995).  Thus, accounting for
$\pm0.15$ dex as a typical uncertainty on the measured iron content, we 
estimate that 
the metallicity of RR Lyr should range from $Z$=0.0005 to $Z$=0.001. 
For the sake of the discussion, let us first adopt the upper limit 
$Z$=0.001.

According to this assumption,  one  can derive 
the stellar mass and luminosity on the basis of evolutionary 
constraints for Horizontal Branch (HB) models. 
Taking into consideration the discrepancies 
still affecting current theoretical
predictions on the ZAHB luminosity  
(see, e.g.,  Castellani 1999; Ferraro et al. 1999; Caputo et al. 2000; 
VandenBerg 2000),  
as well as the increase in luminosity 
expected by  post-ZAHB evolution,  
we end up with log$L/L_{\odot}$=1.70$\pm$0.05. As for the 
stellar mass,  
thanks to the excellent agreement among the various authors about this 
parameter, 
from the evolutionary masses listed in the previous Table 1 we derive  
$M$=0.657$\pm$0.018$M_{\odot}$. We note that current uncertainties 
on the ZAHB luminosity and/or the evolutionary status of RR Lyr introduce  
an uncertainty on $M_K$ of only 0.037 mag, while the uncertainty on 
the mass gives an even smaller error, namely $\sim$ 0.02 mag. 
 
Including these uncertainties in quadrature, and since the 
pulsation period of RR Lyr is log$P$=$-$0.2466 (Hardie 1955), 
the predicted $PLZ_K$ relation 
supplies $M_K$=$-$0.512$\pm$0.045 mag, that  
is well within the error bar of the HST-based estimate 
($M_K$=$-$0.550$\pm$0.118 mag), and with a much lower (formal) 
uncertainty mainly introduced by the uncertainty on the luminosity.
By repeating this evaluation but with $Z$=0.0005, and adopting 
from the relevant literature the evolutionary values  
log$L/L_{\odot}$=1.74$\pm$0.05 and   
$M$=0.697$\pm$0.020$M_{\odot}$, we obtain  
$M_K$=$-$0.570$\pm$0.046 mag. Eventually, 
the predicted $PLZ_K$ relation and evolutionary
constraints provide the following absolute $K$-magnitude for RR Lyr:

$M_K=-0.541\pm 0.044\pm0.029 = -0.541\pm0.062$~mag
  
\noindent
where the final error $\pm$0.062 mag includes the  
additional uncertainty on the adopted metal 
content ($Z$ from 0.0005 to 0.001). 

In conclusion, the predicted   
absolute K-band magnitude of RR Lyr is surprisingly similar to the
HST-based estimate   
($M_K$=$-$0.550$\pm$0.118 mag), and with a formal error that is 
a factor of two smaller than the one introduced by the  
HST absolute parallax.
With $K$=6.54$\pm$0.04 mag, we find that  
the resulting true distance modulus of RR Lyr 
is $\mu_0$=7.081$\pm$0.074 mag, yielding a "pulsation" parallax  
$\pi_{puls}$=3.835$\pm$0.131 mas which is very close to the 
trigonometric parallax measured by HST ($\pi_{abs}$=3.82$\pm$0.20 mas) 
and to the weighted average of HST, Hipparcos, and YPC5 measurements 
($\pi_{abs}$=3.87$\pm$0.19 mas).

The extinction correction only marginally affects this result. 
With the already mentioned estimate $<A_V>$=0.07$\pm$0.03 mag provided by B02, 
the true distance modulus and pulsation parallax 
derived in the absence of reddening 
should be decreased by $\sim$ 0.01 mag and 
increased by $\sim$ 0.01 mas, respectively. 
To be more conservative, we should consider also the
alternative values given by B02 ($<A_V>$=0.11$\pm$0.10 mag) 
and by F98 ($E(B-V)$=0.06$\pm$0.03 mag), adopting  
$<A_V>$=0.12$\pm$0.10 mag as 
a safe estimate of the extinction correction to RR Lyr. 
In such a case, the true distance 
modulus is $\mu_0$=7.068$\pm$0.074 mag and the pulsation parallax 
is $\pi_{puls}$=3.858$\pm$0.131 mas. Once again, the pulsational estimate  
is quite similar to direct trigonometric parallaxes and with a 
total formal error well below the intrinsic uncertainty of the 
HST measurement.

\section{Final remarks}

It has been shown that the predicted $PLZ_K$ relation 
for fundamental RR Lyrae stars   
and current evolutionary constraints on the 
mass and luminosity 
of HB models supply  for RR Lyr itself a "pulsation" parallax   
surprisingly close to the HST measurement 
and to the weighted average of 
HST, Hipparcos and YPC5 measurements, and 
with a better (formal) accuracy 
($\sigma_{\pi}/\pi \sim$ 3.5\%).
We would like to note that without the uncertainty on the adopted metal content 
($0.0005 \le Z \le 0.001$), the error budget on the pulsation parallax 
would be smaller ($\sigma_{\pi}/\pi \sim$ 2\%). Thus, if and when 
the metallicity of RR Lyr will be firmly established, the 
intrinsic uncertainty on the pulsation parallax could be reduced to 
roughly one third of current HST direct measurement ($\sim$ 5\%).

Since the pulsation parallax is well within the error box
of the HST measurement, we can conclude that 
for not heavily reddened RR Lyrae stars the theoretical $PLZ_K$ relation
works {\it at least} as good as HST, almost 
unaffected by uncertainty on the extinction correction. 
Also for  heavily reddened variables, the
pulsation approach could supply distance determinations
as accurate as HST ($\sim$ 5\%),
provided that the extinction correction is known with sufficient accuracy
($\delta A_V\le\pm$0.8 mag).
This means that the $PLZ_K$  
relation 
could be soundly adopted to estimate absolute distances to 
RR Lyrae stars  
in Baade's window as well as in highly reddened regions of the 
Galactic bulge.

However, before claiming that the pulsation approach yields  
distance estimates more precise than trigonometric parallaxes, 
we need further direct  parallax measurements to exclude the 
possible occurrence of (small) systematic errors in the adopted 
(complicated) theoretical scenario. In the meanwhile, 
waiting also for future astrometric missions such as SIM, and 
GAIA, the $PLZ_K$ relation seems actually the only viable means 
to get RR Lyrae distances in the Local Group, as accurate as the 
quoted HST result for the nearby variable RR Lyr.    
 
Finally, we wish to mention that although the HST parallax estimate does 
supply the distance to RR Lyr with an unprecedented accuracy,   
the resulting absolute visual magnitude of this variable cannot firmly 
constrain the zero-point of the 
$M_V$-[Fe/H] relation for RR Lyrae stars, nor discriminate among the 
deceptive discrepancies affecting current theoretical predictions on 
HB evolutionary models, {\it even if the interstellar 
extinction to RR Lyr is firmly known}. As a matter of fact, 
according to the HST parallax (i.e., $\mu_0$=7.090$\pm$0.114 mag) and 
by adopting $V$=7.76 mag (F98) together with the new extinction correction 
$<A_V>$=0.07$\pm$0.03 mag measured by B02, one derives for RR Lyr an 
absolute visual magnitude $M_V$=0.60$\pm$0.118 mag that is still  
affected by a large uncertainty. On the other hand, by adopting this 
extinction correction (i.e. $K_0$=6.532$\pm$0.040 mag) and by taking  
at the face value the predicted near-infrared absolute magnitude 
($M_K$=$-$0.541$\pm$0.062 mag), one would derive 
$\mu_0$=7.073$\pm$0.074 mag and $M_V$=0.617$\pm$0.079 mag. 
On this basis, one can foresee that the application of the predicted 
$PLZ_K$ relation to a large sample of well-studied RR Lyrae stars 
that cover a wide metallicity range and present accurate K-band light 
curves will provide sound constraints on the slope of the 
$M_V$-[Fe/H] relation, as well as on the zero-point of the RR Lyrae 
distance scale if systematic errors can be excluded.

{\bf Acknowledgments:} 
This work was supported by MIUR-Cofin 2000, under the scientific project 
"Stellar Observables of Cosmological Relevance".


\end{document}